Joanna Klauser, Bruno Albert, Christian Lindenmeier, Andreas Hammer, Felix Freiling, Dirk Heckmann, Sabine Pfeiffer:

**Telekommunikationsüberwachung am Scheideweg: Zur Regulierbarkeit des Zugriffes auf verschlüsselte Kommunikation**

*English title and abstract:*

**Telecommunications surveillance at a crossroads: On the regulability of access to encrypted communications**

Personal communication using technical means is protected by telecommunications secrecy. Any interference with this fundamental right requires a legal basis, which has existed for many years for traditional communication services in the form of telecommunications surveillance (TKÜ, § 100a StPO) and appears to be widely accepted by society. The basis for the implementation of TKÜ is the obligation of telecommunications providers to provide interception interfaces. However, the technical implementation of telecommunications has changed significantly as a result of the Internet. Messenger services and Voice over IP telephony are increasingly competing with traditional telephone services. The use of strong end-to-end encryption made possible by this technology is increasingly posing problems for law enforcement agencies, as only cryptographically encrypted content is accessible via the interception interfaces provided to date. Against the backdrop of current discussions on so-called "chat control" and its limited social acceptance, this article addresses the question of whether and, if so, how the cooperation obligations of the technical actors involved can be sensibly regulated in the case of encrypted communication.

*Contact author:*

Felix Freiling, Lehrstuhl für Informatik 1, Friedrich-Alexander-Universität Erlangen-Nürnberg (FAU), Martensstr. 3, 91058 Erlangen, Germany

`felix.freiling@fau.de`

Wiss. Referentin Joanna Klauser, Bayerisches Forschungsinstitut für Digitale Transformation (bidt), München

Wiss. Referent Bruno Albert, Bayerisches Forschungsinstitut für Digitale Transformation (bidt), München

Dr. Christian Lindenmeier, Friedrich-Alexander-Universität Erlangen-Nürnberg (FAU)

Dr. Andreas Hammer, Friedrich-Alexander-Universität Erlangen-Nürnberg (FAU)

Prof. Dr. Felix Freiling, Friedrich-Alexander-Universität Erlangen-Nürnberg (FAU)

Prof. Dr. Dirk Heckmann, Technische Universität München

Prof. Dr. Sabine Pfeiffer, Friedrich-Alexander-Universität Erlangen-Nürnberg (FAU)

# Telekommunikationsüberwachung am Scheideweg

# Zur Regulierbarkeit des Zugriffes auf verschlüsselte Kommunikation

Persönliche Kommunikation mit technischen Mitteln ist durch das Fernmeldegeheimnis geschützt. Eingriffe in dieses Grundrecht bedürfen einer Ermächtigungsgrundlage, die seit vielen Jahren für klassische Kommunikationsdienste in Form der Telekommunikationsüberwachung (TKÜ, § 100a StPO) existiert und gesellschaftlich weitestgehend akzeptiert zu sein scheint. Grundlage für die Umsetzung von TKÜ ist die Verpflichtung der Telekommunikationsanbieter zur Bereitstellung von Abhörschnittstellen. Die technische Umsetzung von Telekommunikation hat sich jedoch durch das Internet stark verändert. Messenger-Dienste und Voice-Over-IP-Telefonie treten zunehmend in Konkurrenz zu klassischen Telefondiensten. Die dadurch ermöglichte Verwendung starker Ende-zu-Ende-Verschlüsselung stellt dabei die Strafverfolgungsbehörden zunehmend vor Probleme, da über die bisher vorgesehenen Abhörschnittstellen nur kryptographisch verschleierte Inhalte zugänglich sind. Vor dem Hintergrund aktueller Diskussionen zur sogenannten „Chatkontrolle" und ihrer eingeschränkten gesellschaftlichen Akzeptanz befasst sich dieser Beitrag mit der Frage, ob und ggf. wie man auch für den Fall verschlüsselter Kommunikation die Mitwirkungspflichten der beteiligten technischen Akteure sinnvoll regulieren kann.

**A. Abhören verschlüsselter Kommunikation gestern und heute**

Die sicherheitspolitische Diskussion über „blinde Flecken" der Strafverfolgung existiert seit Mitte der 1990er Jahre, als verschlüsselte Telekommunikation nur durch Verwendung spezieller Endgeräte möglich war. Die US-Regierung unter Bill Clinton wollte seinerzeit gesetzlich vorschreiben, dass Telekommunikation zwar standardmäßig verschlüsselt werden sollte, dass jedoch gleichzeitig alle Endgeräte durch den Einbau des so genannten „Clipper-Chips" für jede verschlüsselte Kommunikationsverbindung einen Nachschlüssel für staatliche Behörden (Key Escrow) vorhalten mussten. Dies führte zu erhitzten gesellschaftlichen Diskussionen, die auch durch Kritik führender Wissenschaftler und Wissenschaftlerinnen aus dem Bereich der Informatik befeuert wurden.[1]

Am Ende wurde die Regelung aufgegeben, wohl auch, weil damals bereits ein Großteil des weltweiten Datenverkehrs über das neue „Internet" abgewickelt wurde, und zwar in unverschlüsselter Form. Aber auch dort verfestigte sich der Trend hin zu verschlüsselter Kommunikation, spätestens seit der Standardisierung des Internet-Verschlüsselungsprotokolls TLS und der allgemeinen Unterstützung von HTTPS durch Webbrowser um die Jahrtausendwende.

Die Diskussion über staatliche Zugänge zu verschlüsselter Kommunikation riss jedoch trotz der Enthüllung überbordender Abhörpraktiken der NSA durch Edward Snowden im Jahr 2013 nicht ab.[2] In Deutschland wurden bereits 2006 Fälle bekannt,[3] bei denen Sicherheitsbehörden teilweise ohne explizite Ermächtigungsgrundlagen mit verdeckten technischen Mitteln in Endgeräte eingriffen, um auf verschlüsselte Kommunikation zuzugreifen. In Anlehnung an die klassische TKÜ wird dieses Vorgehen als „Telekommunikationsüberwachung an der Quelle" (Quellen-TKÜ, QTKÜ) bezeichnet. Die darauffolgende Entscheidung des Bundesverfassungsgerichts zur Online-Durchsuchung[4] erschuf nicht nur ein neues Grundrecht,[5] sondern formulierte auch Leitplanken für die verfassungsgemäße Ausgestaltung derartiger Abhörvorgänge auf persönlichen Endgeräten. Den Leitplanken dieser Entscheidung folgen die heute gültigen Ermächtigungsgrundlagen in der StPO, die wir im Abschnitt 3 näher erläutern.

Die sicherheitspolitische Diskussion dauert bis heute an und wurde zuletzt neu belebt durch den Verordnungsentwurf 2022/0155 (COD) der EU-Kommission zur Bekämpfung und Verhinderung von sexuellem Kindesmissbrauch im Internet.[6] Über das Vorhaben der Chatkontrolle konnte bisher trotz zahlreicher Versuche keine Einigung erzielt werden.[7] Der Verordnungsentwurf 2022/0155 (COD) der EU-Kommission sah vor, dass die Anbieter von Hosting- oder Messenger-Diensten dazu verpflichtet werden, eine Risikobewertung vorzunehmen, inwieweit ihre Dienste für die Verbreitung von Material über sexuellen Kindesmissbrauch verwendet werden. Zudem waren auch Risikominderungspflichten vorgesehen bis hin zur automatisierten Prüfung von Inhalten auf potenzielle Strafbarkeit.[8]

---

[1] *Abelson/Anderson/Bellovin/Benaloh/Blaze/Diffie/Gilmore/Neumann/Rivest/Schiller/Schneier* World Wide Web Journal 1997, 241. Neben der grundsätzlichen Kritik an der vorgeschriebenen Schwächung von Kryptographie wurde auch argumentiert, dass entsprechend ausgestattete Geräte teurer und deshalb unattraktiver für Kunden würden und dies gerade US-amerikanische Hersteller benachteiligte.

[2] *Abelson/Anderson/Bellovin/Benaloh/Blaze/Diffie/Gilmore/Green/Landau/Neumann/Rivest/Schiller/Schneier/Specter/Weitzner* Journal of Cybersecurity 2015, 69.

[3] BGH MMR 2007, 174; BGHSt 51, 211.

[4] BVerfGE 120, 274.

[5] *Hoffman-Riem* JZ 2008, 1009 (1009).

[6] *Abelson/Anderson/Bellovin/Benaloh/Blaze/Callas/Diffie/Landau/Neumann/Rivest/Schiller/Schneier/Teague/Troncoso* Journal of Cybersecurity 10 Ausgabe 1, 1.

[7] *Meister*, Polen gibt Einigung bei Chatkontrolle auf, Netzpolitik.org, Stand: 26.06.2025, https://netzpolitik.org/2025/interne-dokumente-polen-gibt-einigung-bei-chatkontrolle-auf/.

**B. Aktuelle Ermächtigungsgrundlagen in der StPO**

Demgegenüber dient sowohl die klassische TKÜ als auch die QTKÜ der Erfassung der Telekommunikation von Personen zu Zwecken der Strafverfolgung.[9] Bei der Durchführung der Maßnahmen ist die Gebietshoheit der betroffenen Staaten zu wahren. Diese besagt, dass deutsche Strafverfolgungsbehörden nur auf deutschem Staatsgebiet ermitteln dürfen.[10] Telekommunikation findet indessen auch über Grenzen hinweg statt. Dementsprechend regelt § 4 TKÜV auch die Zulässigkeit der Überwachung von Verbindungen zwischen Anschlüssen im Inland mit solchen im Ausland. Die völkerrechtliche Zulässigkeit der Norm wird damit begründet, dass durch den Zugriff an den „Gateways" der Inlandsverbindung die Maßnahme auf deutschem Staatsgebiet stattfindet.[11] Die Argumentation lässt sich insofern auf die QTKÜ übertragen, als die Überwachung der Kommunikation nur durch Zugriffe auf Geräte, welche sich im Inland befinden, erfolgt. Der Zugriff auf Daten, welche auf im Ausland verorteten Geräten gespeichert sind, ist im Grundsatz völkerrechtswidrig.[12] Bezüglich der völkerrechtlichen Bewertung bei erschwerter oder unmöglicher Ermittlung des Standorts herrscht Uneinigkeit.[13] Dies darf indes nicht als Ausrede verwendet werden, die bestehenden Möglichkeiten nicht erschöpfend zu beleuchten.[14]

TKÜ und QTKÜ unterliegen den gleichen Anforderungen an den Eingriff (§§ 100a Abs. 1 S. 1, Abs. 2 [15], 3, 100d Abs. 1 StPO), die Anordnung und die Anordnungsdauer (§ 100e Abs. 1, 3, 4, 5 StPO). Trotz dieser gemeinsamen Normierung unterscheiden sich die Befugnisse hinsichtlich ihrer technischen Umsetzung allerdings erheblich.

---

[8] Aufgrund einer Aufdeckungsanordnung auf Antrag einer Koordinierungsbehörde, bestätigt durch ein Gericht oder eine unabhängige Verwaltungsstelle bestünde für Hosting-Anbieter und interpersonelle Kommunikationsdienste nach dem ersten Entwurf die Pflicht mittels „Technologien" kinderpornographische Abbildungen und Grooming zu detektieren (Art. 7 Abs. 1, Art. 10 Abs. 1 VO-Entwurf 2022/0155(COD). Während der Vorschlag die genaue Technik zur Inhaltsprüfung nicht präzisierte (in Art. 10 Abs. 3 VO-Entwurf 2022/0155(COD) finden sich lediglich einige Zielvorgaben; EG 26 erläutert, dass die Technologie, mittels welcher die Aufdeckungsanordnungen ausgeführt werden, durch die Anbieter auszuwählen ist), nimmt er Anbieter von Ende-zu Ende verschlüsselter Kommunikation nicht aus und wirft damit die Frage der Umsetzung auf.

[9] Im Folgenden liegt der Fokus auf den Ermächtigungsgrundlagen der StPO, auch wenn funktionsgleiche Eingriffsermächtigungen im Gefahrenabwehrrecht existieren, welche sowohl die TKÜ als auch die QTKÜ in technischer Hinsicht ähnlich bis deckungsgleich normieren: TKÜ: Art. 51 Abs. 1 BKAG, Bsp. des Landesrechts: Art. 42 Abs. 1 BayPAG; QTKÜ: Art. 51 Abs. 2, 1 BKAG, Bsp. des Landesrechts: Art. 42 Abs. 2, 1 BayPAG.

[10] *Arnauld*, Völkerrecht, 5. Aufl. 2023, § 4 C I 1 Rn. 341; *Epping*, in: Ipsen (Hrsg.), Völkerrecht, 8. Aufl. 2024, § 7 Rn. 69.

[11] Siehe *Tiedemann* CR 2005, 858 (862); KMR-StPO/*Bär*, 133.Lfg., §100a Rn. 82; i.E. auch MüKo-StPO/*Rückert*, 2. Aufl. 2023, § 100a Rn. 250; BeckOK-StPO/*Graf*, 55. Ed. 1.1.2025, § 100a Rn. 280; Unter Bezugnahme auf fehlende physikalisch-technische Wirkung der Überwachung im fremden Hoheitsgebiet, siehe *Kilching*, Die Neuregelung zur Auslandskopfüberwachung gemäß § 4 TKÜV auf dem verfassungsrechtlichen Prüfstand, 2006, S. 30; Für Geheimdienstliche Telekommunikation: *Weber/Sravia* NJW 2007, 1433 (1435); a.A. Unterscheidung zwischen der Überwachung ausländischer Anschlüsse um Kommunikation mit inländischen Anschlüssen abzufangen und der inländischer Anschlüsse um Kommunikation mit ausländischen Anschlüssen abzufangen: *Vehling*, Die Auswirkungen des Völkerrechts auf die grenzüberschreitende Ermittlung digitaler Beweise nach der StPO, 2023, S. 126 ff.; Unter Bezugnahme auf die Auswirkung der Maßnahme im Ausland, *Reinel* wistra 2006, 205 (207).

[12] Tallin Manual 2.0, 2017 Rule 11 Nr. 14.; *Vehling*, Auswirkungen des Völkerrechts auf grenzüberschreitende Ermittlung digitaler Beweise (Fn. 11), S. 129 ff.; Rechtswidrigkeit der räumlichen Erstreckung von Durchsuchungen auf Speichermedien im Ausland gemäß § 110 Abs. 3 StPO, siehe BeckOK-IT-Recht/*Brodowski*, 18. Ed. 2024, StPO § 110 Rn. 12; *Babuke/Kroner* NZWiSt 2024, 174 (179 f.). Allerdings kann die Zustimmung des betroffenen Staates die völkerrechtliche Zulässigkeit der Maßnahme bewirken. Siehe Tallin Manual 2.0, Rule 11 (b), Nr. 7. Möglich ist neben der fallbezogenen Zustimmung auch die Anwendbarkeit eines Abkommens auf den Fall, wie z. B. in bestimmten Fällen durch die Cybercrime Convention, siehe Tallin Manual 2.0, Rule 11, 9; *Brodowski*, in: Hauck/Peterke (Hrsg.), International law and transnational organized crime, 2016, S. 355f.

[13] Tallin Manual 2.0 Rule 11 Nr. 8.

[14] So aber LG Koblenz zu § 110 Abs. 3 StPO: LG Koblenz NZWiSt 2022, 160; kritisch *Bechtel* NZWiSt 2022, 160 (162 f.).

[15] Hinsichtlich des Straftatenkatalogs nach dem Urteil des BVerfG vom 24.06.2025 ist die Befugnis zur QTKÜ in § 100 Abs. 2 StPO in Bezug auf einige Straftaten nichtig. Die QTKÜ begründet, im Gegensatz zur TKÜ nicht nur einen Eingriff in das Kommunikationsgrundrecht, sondern muss sich auch am IT-System-Grundrecht messen lassen. BVerfG BeckRS 2025, 19413 - *Trojaner II*.

Um die klassische TKÜ gemäß § 100a Abs. 1 S. 1 StPO zu ermöglichen, werden die Betreiber von Telekommunikationsanlagen gemäß § 100a Abs. 4 S. 2 StPO i.V.m. § 170 Abs. 1 S. 1 Nr. 1 TKG verpflichtet „auf eigene Kosten technische Einrichtungen zur Umsetzung gesetzlich vorgesehener Maßnahmen zur Überwachung der Telekommunikation vorzuhalten und organisatorische Vorkehrungen für deren unverzüglichen Umsetzung zu treffen". Weitere Bestimmungen zu der Maßnahmenimplementierung finden sich in § 170 TKG, der TKÜV (Verordnung über die technische und organisatorische Umsetzung von Maßnahmen zur Überwachung der Telekommunikation) und der TR TKÜV (technische Richtlinie zur Umsetzung gesetzlicher Maßnahmen zur Überwachung der Telekommunikation, Erteilung von Auskünften, Ausgabe 8.3). Erstere wird von der Bundesregierung mit Zustimmung des Bundesrates erlassen und präzisiert u.a. grundlegende technische Anforderungen und organisatorische Eckpunkte (§ 170 Abs. 5 Nr. 1 lit. a TKG): In der TKÜV wird festgelegt, dass die Überwachung gemäß § 5 Abs. 2 TKÜV ausgeführt wird, indem die Verpflichteten (§ 2 Nr. 16 TKÜV) an einem Übergabepunkt (§ 2 Nr. 11 TKÜV) eine Kopie der Telekommunikation (§ 2 Nr. 14 TKÜV) für die berechtigte Stelle (§ 2 Nr. 3 TKÜV) bereitstellen. Sie enthält darüber hinaus auch Anforderungen an diesen Prozess betreffend den Schutz gegen unbefugte Inanspruchnahme (§ 14 Abs. 1 TKÜV) und gegen Kenntnisnahme durch unbefugte Dritte (§ 14 Abs. 2 TKÜV) sowie an die Dokumentation des Vorgangs (§ 16 TKÜV). In der TR TKÜV werden die technischen Details durch die Bundesnetzagentur unter Einbindung der berechtigten Stellen, Verbände und Hersteller und unter Berücksichtigung internationaler technischer Standards normiert (§ 170 Abs. 6 TKG, § 36 TKÜV). Der Bundesnetzagentur kommt außerdem eine Prüffunktion zu. Denn die Betreiber von Telekommunikationsanlagen haben ihr gemäß § 170 Abs. 1 Nr. 4 TKG den Nachweis zu erbringen, dass ihre technischen Einrichtungen und organisatorischen Vorkehrungen den Anforderungen aus der TKÜV und TR TKÜV entsprechen. Während die Gerichte als rechtliche Kontrollinstanz der Maßnahme fungieren (§ 100e Abs. 1 StPO), können auch die verpflichteten Telekommunikationsdienstleister durch ihre Einbindung in die Durchführung einen Missbrauch verhindern [16].

---

[16] Nach § 12 Abs. 2 S. 1 TKÜV setzt der Verpflichtete die gerichtliche Anordnung um, nachdem er diese auf gesicherten elektronischen Weg oder eine Kopie per Fax erhalten hat. Erhält er gemäß § 12 Abs. 2 S. 2 TKÜV nicht binnen einer Woche das Original, muss er die Vorrichtung abschalten. Darüber hinaus wird den Telekommunikationsdienstleistern kein Prüfungsrecht hinsichtlich der Rechtmäßigkeit der Maßnahme zugesprochen (BeckOK-StPO/ *Graf*, 56. Ed. 2025, § 100a Rn. 175, 273, KK-STPO/ *Heinrichs/Weingast*, 9. Aufl. 2023, § 100a Rn. 37; BGH Ermittlungsrichter MMR 1999, 99 (100). Die Diensteanbieter könnten nicht die Rechte der Betroffenen verteidigen und ein solches Recht würde die effektive Umsetzung der Maßnahme verhindern (BGH, Ermittlungsrichter MMR 1999, 99, (100). Teilweise wird aber eine Prüfung formeller Voraussetzungen (*Wolf/Neumann* NstZ 2003, 404 (407) bzw. eine Art Willkürprüfung (KK-STPO/*Heinrichs/Weingast*, 9. Auf. 2023, § 100a Rn. 37; wohl auch BeckOK-StPO/*Graf*, 56 Ed. 2023, § 100a Rn. 272) für zulässig erachtet. Festhalten lässt sich, dass jedenfalls ein starker Missbrauch seitens des Staates durch Telekommunikationsanbieter auffallen und verhindert werden könnte. Vgl. *Schwabenbauer*, Heimliche Grundrechtseingriffe, 2013, S. 178; *Rückert*, Digitale Daten als Beweismittel im Strafverfahren, 2023, S. 281.
Gleichzeitig werden die verpflichteten Telekommunikationsdienstleister selbst kontrolliert: Sie müssen u.a. eine Protokollierung der Zeitpunkte des Zugriffs gemäß § 16 Abs. 1 Nr. 3 TKÜV vornehmen und einen Teil der Protokolle gemäß § 17 Abs. 1 S. 1 TKÜV auf Übereinstimmung mit den ihnen vorliegenden Unterlagen prüfen. Eine Kopie der Prüfergebnisse ist gemäß § 17 Abs. 9 TKÜV vierteljährlich an die Bundesnetzagentur zu senden. Gemäß § 17 Abs. 3 S. 1 TKÜV muss der Verpflichtete bei Beanstandungen unverzüglich eine Untersuchung einleiten und die Bundesnetzagentur unterrichten. Die Prüfpflichten sollen einem Missbrauch der technischen Funktionen aufdecken (Beck'scher TKG-Kommentar/*Eckhardt*, 5. Aufl. 2023, § 170 Rn. 99).

Die Kommunikationsüberwachung auf klassischem Wege ist bei Ende-zu-Ende Verschlüsselung nicht möglich, weil die Betreiber von Telekommunikationsanlagen die Verschlüsselung der Kommunikation nicht aufzuheben vermögen. Eine etwaige Pflicht zur Vorhaltung eines Schlüssels ergibt sich insbesondere nicht aus der in § 100a Abs. 4 StPO normierten Mitwirkungsverpflichtung.[17] Die QTKÜ ist deshalb im Unterschied zu der klassischen TKÜ bisher ohne Einbindung dritter Parteien reguliert. § 100a Abs. 1 S. 2, 1 StPO ermöglicht den Strafverfolgungsbehörden eine Überwachung von verschlüsselter Kommunikation durch den Eingriff in informationstechnische Systeme „mit technischen Mitteln". Zwar stellt das Gesetz in § 100a Abs. 5 StPO bestimmte Anforderungen an die Mittel, welche zum Teil den Anforderungen in der TKÜV ähneln,[18] so wird in § 100a Abs. 5 Nr. 1 lit. a StPO eine technische Begrenzung der Maßnahme auf die laufende Telekommunikation gefordert bzw. gemäß § 100a Abs. 5 Nr. 1 lit. b StPO auf Inhalte und Umstände der Kommunikation, die ab dem Zeitpunkt der Anordnung während des laufenden Übertragungsvorgangs im öffentlichen Telekommunikationsnetz hätten überwacht und aufgezeichnet werden können. Die genauere Ausgestaltung des Eingriffs und des technischen Mittels bleibt aber offen. Und genau dies sorgt in der Umsetzung für Probleme.

## C. Technische Optionen des Zugriffs

Die bemerkenswerte „Nicht-Regulierung" der Umstände einer QTKÜ hat zu intensiven Diskussionen in der rechtswissenschaftlichen Literatur geführt über die Frage, was als „technisches Mittel" im Sinne des § 100a Abs. 1 S. 2, 1 StPO zu verstehen ist. Klar erfasst erscheinen der Einsatz einer *Remote Forensic Software* (umgangssprachlich „Staatstrojaner") sowie verschiedene Hardwarelösungen[19]. Des Weiteren sind verschiedene Möglichkeiten denkbar, um auf das von dem Betroffenen genutzte System zuzugreifen. Es können Schwachstellen genutzt werden, um aus der Ferne Zugriff zu erlangen[20] oder eine Software wird im Rahmen eines kurzzeitigen physischen Zugriffs auf das Gerät aufgespielt[21]. Ebenfalls diskutiert wird der Zugang durch Täuschung und Mitwirkung des Betroffenen[22] sowie die

---

[17] BeckOK IT-Recht/*Brodowski*, 17. Ed. 2024, StPO § 100a Rn. 26. Etwas anderes gilt für den Anbietern bekannte Schlüssel gemäß § 7 Abs. 1 S. 1 Nr. 10 TKÜV. Vgl. MüKo-StPO/*Rückert*, 2. Aufl. 2023, StPO § 100a Rn. 255.

[18] Gesetzlich festgelegt wird u. a., dass technisch sicherzustellen sei, dass nur unerlässliche Veränderungen am System vorgenommen würden (§ 100a Abs. 5 S. 1 Nr. 2 StPO), welche möglichst rückgängig zu machen sind (§ 100a Abs. 5 S. 1 Nr. 3 StPO) und, dass das Mittel nach Stand der Technik gegen unbefugte Nutzung (§ 100a Abs. 5 S. 2 StPO) und die kopierten Daten gegen Veränderung, unbefugte Löschung und Kenntnisnahme (§ 100a Abs. 5 S. 3 StPO) zu schützen seien. Des Weiteren wurden Protokollierungspflichten normiert (§ 100a Abs. 6 StPO). Die gesetzlichen Anforderungen an das Mittel sollen durch die Standardisierte Leistungsbeschreibung genauer dargestellt werden, wenngleich diese schon im Umfang wesentlich weniger detailliert ist, als die TKÜV RL. (Standardisierende Leistungsbeschreibung für Software zur Durchführung von Maßnahmen der Quellen-TKÜ und Online-Dursuchung, BKA, Stand: 13.02.2025, https://www.bka.de/DE/UnsereAufgaben/Ermittlungsunterstuetzung/Technologien/QuellentkueOnlinedurchsuchung/quellentkueOnlinedurchsuchung_node.html).

[19] BeckOK IT-Recht/*Brodowski*, 17. Ed. 2024, StPO § 100a Rn. 8, 100b Rn. 6; BeckOK-StPO/*Graf*, 54. Ed. 2025, § 100a Rn. 126; KK-StPO/*Henrichs/Weingast*, 9. Aufl. 2023, § 100a Rn. 42; MüKoStPO/*Rückert*, 2. Aufl. 2023, § 100a Rn. 203; ablehnend bzgl. Seitenkanalattacken i. R. v. § 100a StPO vgl. *Seum/Krüger/Wildermann/Teich*, MMR 2025, 98 (101 f.).

[20] MüKoStPO/*Rückert*, 2. Aufl. 2023, § 100a Rn. 204; kritisch bzgl. bewussten Offenhalten von IT-Sicherheitslücken durch Behörden: *Derin/Golla* NJW 2019, 1111 (1114 f.).

[21] BeckOK IT-Recht/*Brodowski*, 17. Ed. 2024, StPO § 100a Rn. 8, § 100b Rn. 7; BeckOK-StPO/*Graf*, 54. Ed. 2025, § 100a Rn. 132.

[22] Befürwortend BeckOK-StPO/*Graf*, 54. Ed. 2025, § 100a Rn. 129; KK-StPO/*Henrichs/Weingast*, 9. Aufl. 2023, § 100a Rn. 46; MüKo-StPO/*Rückert*, 2. Aufl. 2023, § 100a Rn. 207; kritisch *Derin/Golla* NJW 2019, 1111 (1113 f.).

Verwendung eines anderweitig erlangten Passworts[23]. Das heimliche Betreten von Wohnraum, um einen Zugriff auf Endgeräte zu erhalten, ist nicht erfasst.[24]

Die unübersichtliche Vielzahl an Zugriffsoptionen ist eine direkte Folge der Unbestimmtheit der rechtlichen Norm und der politischen Entscheidung des Gesetzgebers, Mitwirkungspflichten von Dritten nicht in Betracht zu ziehen. Wir versuchen deshalb im Folgenden, dieses Feld auf Basis von technischen Kriterien zu systematisieren, um die grundsätzlichen Schwierigkeiten seiner Regulierung herauszuarbeiten.

### I. Der Raum möglicher Zugriffsoptionen

Die grundsätzlichen Optionen des technischen Zugriffs auf verschlüsselte Kommunikation lassen sich generell anhand der Annahme über den *Zugang* zum abzuhörenden Endgerät unterscheiden. Der Begriff „Zugang" bezieht sich hierbei auf die Möglichkeit, Software mit Abhörfunktionalität auf dem Zielsystem auszuführen. Dies ist zentral, da bei Ende-zu-Ende-Verschlüsselung definitionsgemäß nur dort die Kommunikationsdaten in unverschlüsselter Form vorliegen. Ohne einen solchen Zugang bleibt den Strafverfolgungsbehörden grundsätzlich nur der Weg über die klassische TKÜ.[25] Dies ist in Abbildung 1 dargestellt.

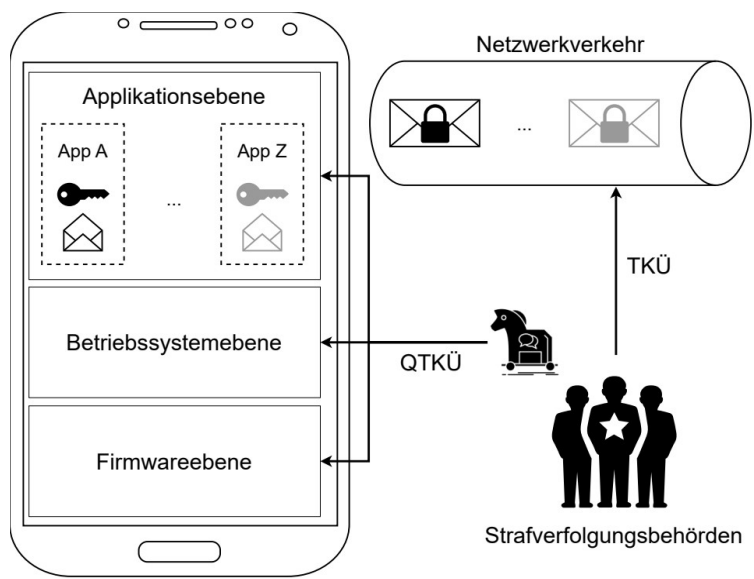

*Abbildung 1: Die Strafverfolgungsbehörden haben die Optionen einer TKÜ und QTKÜ. Während Kommunikationsdaten die mit einer TKÜ aufgezeichnet werden verschlüsselt sind, liegen die unverschlüsselten Daten und die zugehörigen Schlüssel auf dem Endgerät. Bei der QTKÜ wird eine forensische Software („Trojaner") auf einer der Privilegienstufe installiert.*

Ähnlich den Diskussionen um den Clipper-Chip in den 1990er Jahren, gibt es immer wieder Vorschläge, kryptographische Protokolle und Algorithmen derart zu modifizieren, dass

---

[23] MüKo-StPO/*Rückert*, 2. Aufl. 2023, § 100a Rn. 203.
[24] BeckOK-StPO/*Graf*, 54. Ed. 2025, § 100a Rn. 129; BeckOK IT-Recht/*Brodowski*, 17. Ed. 2024, StPO § 100a Rn. 8, § 100b Rn. 7; KK-StPO/*Henrichs/Weingast*, 9. Aufl. 2023, § 100a Rn. 46; MüKo-StPO/*Rückert*, 2. Aufl. 2023, § 100a Rn. 205.
[25] Auch wenn es Ansätze gibt, die versuchen anhand von z.B. Metadaten Rückschlüsse auf die Kommunikationsinhalte zu ziehen, sind diese jedoch oftmals unzureichend und unzuverlässig, siehe *Velan/Čermák/Čeleda/Drašar* International Journal of Network Management 2015, 355.

Strafverfolgungsbehörden eine Art „Masterschlüssel" besitzen, der für die Entschlüsselung genutzt werden kann.[26] Wie eingangs berichtet, stoßen diese Ansätze jedoch regelmäßig auf starke öffentliche Kritik. Sie sind aber auch technisch überaus komplex, da die Voraussetzung für derartige Ansätze der Betrieb einer Infrastruktur für das sichere Verwahren von kryptographischen Schlüsseln ist.[27] Während diese Option durchaus auch potentielle Vorteile bieten kann (z.B. garantierte Begrenzung auf die laufende Kommunikation, private Daten auf dem Endgerät sind vom Zugriff ausgeschlossen), gilt die sichere Umsetzung eines solchen Konzepts heute als nicht sinnvoll umsetzbar.[28]

**II. Zugang zum Endgerät auf verschiedenen Privilegienstufen**

Digitale Endgeräte sind komplexe informationstechnische Systeme, die aus unterschiedlichen Sicherheitszonen bestehen, in denen Software nur dann operieren kann, wenn diese die richtigen Privilegien hat. Dies ist analog zu den Freigabestufen von Verschlusssachen in Behörden (vertraulich, geheim, streng geheim), bei denen beispielsweise eine Person mit Freigabestufe „geheim" keine Dokumente lesen darf, die als „streng geheim" eingestuft sind.

Typisch für moderne mobile Geräte (z.B. Android Smartphones) sind drei Privilegienstufen: Applikationsebene, Betriebssystemebene und Firmwareebene (siehe Abbildung 1). Generell ist ein Zugriff auf Daten einer Privilegienstufe nur von derselben oder einer höher liegenden Privilegienstufe möglich.[29] Auf der *Applikationsebene* laufen Apps wie z.B. WhatsApp, Signal oder Telegram, die jedoch voneinander isoliert sind. Auf der *Betriebssystemebene* ist z.B. Android oder iOS installiert, welches Zugriff auf die Daten aller Apps hat. Die am höchsten privilegierte Ebene – die *Firmwareebene* – beinhaltet Software, die typischerweise direkt vom Geräte- oder Chiphersteller (z.B. Samsung, Apple oder Qualcomm) bereitgestellt wird und nicht vom Nutzer modifiziert werden kann.

Die Privilegienstufe des Zugangs, also der forensischen Software, mit der die Strafverfolgungsbehörden eine QTKÜ auf dem Endgerät durchführen, ist ein wesentliches technisches Kriterium, um verschiedene Möglichkeiten des Zugriffs auf verschlüsselte Kommunikation zu unterscheiden. Ein zweites, davon unabhängiges Kriterium betrachtet den Weg, wie dieser Zugang erlangt wird. Aus technischer Sicht gibt es schlussendlich lediglich zwei universelle Optionen: Entweder können Schwachstellen ausgenutzt werden oder es findet eine Kooperation mit dem Anbieter der Software auf einer der drei Privilegienstufen statt. Die Unterscheidung dieser beiden Optionen ist bei der Bewertung möglicher Regulierungsoptionen wichtig.

**III. Zugang durch Ausnutzen von Schwachstellen**

---

Die wohl bekannteste Vorgehensweise zur Herstellung des Zugangs zum Endgerät ist das Ausnutzen von Schwachstellen mithilfe von sogenannten „Exploits". Dabei werden – typischerweise öffentlich unbekannte – Schwachstellen durch das Versenden von schadhaften Daten ausgenutzt, um eigene Software auf dem Zielgerät zu installieren. Die dabei erreichte Privilegienstufe hat auch hier starken Einfluss auf die Umstände der QTKÜ. So ermöglicht ein Exploit auf Applikationsebene normalerweise nur Zugriff auf Daten einer spezifischen App, während ein Exploit gegen das Betriebssystem dazu führt, dass die Daten aller Apps zugreifbar werden.

Das Nutzen von Exploits hat durchaus Vorteile gegenüber anderen Optionen. Zum einen ist die Herstellung des Zugangs eine zielgerichtete und einmalige Aktion, sodass zunächst nur genau ein Endgerät betroffen ist. Außerdem sind Exploits eine Universallösung, denn sie funktionieren – mit genug Aufwand und der Annahme, dass kein System frei von Schwachstellen ist – auf beliebigen Geräten. Auch der Umstand, dass Exploits technisch komplex, und dadurch selten sind, kann als „natürliche Begrenzung" des Umfangs staatlicher Überwachung angesehen werden.

Trotzdem haben Exploit-basierte Zugangsoptionen eine Reihe von Nachteilen. Zum einen kann argumentiert werden, dass sie keine nachhaltige und zuverlässige Lösung darstellen, da z.B. oftmals für neuere Geräte keine Exploits verfügbar sind. Die hohen Aufwände zur Entwicklung von Exploits machen diese kostbar und ökonomisch wertvoll. So ist ein Markt privater Anbieter entstanden, die nicht nur staatliche Stellen mit Zugängen versorgen.[30] Gleichzeitig ist selbst das Zukaufen von Exploits durch legitime staatliche Behörden problematisch, denn die Intransparenz der gekauften technischen Mittel führt zu einer niedrigen forensischen Beweiskraft der so erhobenen Daten.[31] Die Problematik der Unsicherheit ist jedoch nicht vollständig zu lösen, selbst wenn die staatlichen Behörden ihre Exploits eigenständig entwickeln würden. Denn sobald Ermittler einmalig Vollzugriff auf ein Endgerät erhalten (d.h. mindestens auf Betriebssystemebene), sind Manipulationen technisch nicht mehr auszuschließen. Während das reine Nutzen von Exploits auf Applikationsebene hier eine Teillösung verspricht, scheitert diese jedoch oft daran, dass es eine Vielzahl an potentiellen Apps gibt, sodass die Verfügbarkeit eines passenden Exploits typischerweise nicht gegeben ist.

Ein weiterer Hauptkritikpunkt von exploit-basierten Zugängen ist das entstehende Spannungsfeld zwischen IT-Sicherheitszielen und staatlichem Zugang.[32] Da Exploits eine meist unbekannte Schwachstelle ausnutzen, darf diese – zumindest vorübergehend – nicht geschlossen werden. Das führt jedoch dazu, dass staatliche Akteure wissend in Kauf nehmen, dass diese Schwachstellen auch von Kriminellen ausgenutzt werden könnten.[33]

---

[30] *Kaster/Ensign* Thunder-bird International Business Review 65, 355; auch: *Perlroth*, This Is How They Tell Me The World Ends, 2021.
[31] *Ottmann/Pollach/Scheler/Schneider/Rückert/Freiling* DuD 2021, 546 (548).
[32] BVerfG BeckRS 2025, 19413 - *Trojaner II*, Rn. 194; *Wagner/Vettermann/Arzt/Brodowski/Dickmann/Golla/Goerke/Kreutzer/Leicht/Obermaier/Schink/Schreiber/Sorge*, Verantwortungsbewusster Umgang mit IT-Sicherheitslücken: Problemlagen und Optimierungsoptionen für ein effizientes Zusammenwirken zwischen IT-Sicherheitsforschung und IT-Verantwortlichen, 2016, (30 ff.).

**IV. Zugang auf Basis von Herstellerkooperation**

Während technische Unterstützung für die Durchführung der TKÜ durch die Telekommunikationsanbieter gegeben ist, sind Hersteller von Endgeräten oder deren Software – wie oben beschrieben – bisher nicht zur Unterstützung verpflichtet. Die Verpflichtung dieser Parteien erscheint vor dem Hintergrund der Erfahrungen mit klassischer TKÜ verlockend, aber auch hier gibt es Vor- und Nachteile.

So sind bei den weitverbreitetsten Endgeräten regelmäßig eine Vielzahl von Herstellern auf unterschiedlichen Privilegienstufen involviert.[34] Eine Verpflichtung der Hersteller auf Applikationsebene hätte zur Folge, dass Anbieter von verschlüsselter Kommunikation (z.B. WhatsApp, Signal, Threema oder Telegram) nach Anfrage durch die Strafverfolgungsbehörden gezielt eine Abhörschnittstelle auf das entsprechend Endgerät bereitstellen müssten. Dadurch könnten Kommunikationsdaten vor oder nach der Entschlüsselung im Speicher abgefangen werden. Das Hinzufügen einer solchen Abhörschnittstelle durch den Softwarehersteller ist technisch zwar aufwendig, jedoch weniger komplex, und damit argumentativ zuverlässiger, als eine Exploit-basierte Umsetzung, da der Hersteller den Quellcode der App anpassen kann. Damit geht auch der Vorteil einher, dass hier gezielter nur genau die Daten ausgeleitet werden können, die auch für die Ermittlungen relevant sind, da kein Vollzugriff auf das Gerät besteht. Im Hinblick auf Manipulationen ist jedoch die Applikationsebene ungünstig, da die abgefangenen Daten jeweils durch höhere Privilegienstufen verarbeitet werden und durch diese manipuliert werden könnten.

Während auf Applikationsebene eine Vielzahl an Herstellern verpflichtet werden müssten, ist der Umstand auf der Betriebssystemebene anders. Hier gibt es weitaus weniger Anbieter (z.B. Google mit Android und Apple mit iOS). Die höhere Privilegienstufe ermöglicht hier zwar auch den Zugriff auf die Kommunikationsdaten in unverschlüsselter Form aller Apps, jedoch erfordert der Zugriff eine komplexere Programmlogik.[35] Ungünstig ist außerdem auf dieser Privilegienstufe, dass die Strafverfolgungsbehörden Zugriff auf beinahe alle Daten des Endgerätes erlangen, wodurch ein erhöhtes Missbrauchspotential entsteht. Und auch auf der Betriebssystemebene bleibt das Restrisiko von Manipulationen durch Software auf der höher liegenden Firmwareebene bestehen.

---

[33] Zur Frage des Tätigwerdens der Polizei zur Gewährleistung der inneren Sicherheit siehe BVerfG NJW 2021, 3033 Ls. 2, Rn. 26, 33; ausführlich diskutiert von *Gruber/Brodowski/Freiling* GSZ 2022, 171.

[34] Beispiel iPhone: Apple und der App-Hersteller (Whatsapp-Meta, Signal, Telegram). Beispiel Android: Gerätehersteller (z.B. Samsung), OS-Hersteller (Google) und der App-Hersteller. Wir vernachlässigen im Folgenden die Möglichkeit, den Hardware-Hersteller zum Einbau einer Abhörschnittstelle analog zum Clipper-Chip zu verpflichten, da zu viele Fragen aus technischer Sicht ungeklärt sind. So ist der Betrieb eines solchen Chips bereits technisch anspruchsvoll und es müsste sichergestellt sein, dass regelmäßige Updates geliefert werden, sodass die Funktionsweise trotz technischen Fortschritts gesichert ist. Außerdem müsste eine solch vorinstallierte Hintertür zudem vor Manipulationen und dem Entfernen geschützt werden, ein technisch ungelöstes Problem.

[35] Die Abhörsoftware müsste pro Kommunikationsapp speziell konfiguriert werden, um die Nachrichten in unverschlüsselter Form im Speicher zu finden.

Wenn der Zugang auf der höchsten Privilegierungsstufe, also auf Firmwareebene, durch Verpflichtung der Gerätehersteller erfolgt, bestehen keine Bedenken hinsichtlich Manipulation durch Dritte. Auch können auf dieser Stufe spezielle Hardwareerweiterungen (z.B. ARM's TrustZone oder Apple's Secure Enclave) verwendet werden, um sicherzustellen, dass nur vertrauenswürdige, d.h. vom Gerätehersteller signierte, Software auf dem Gerät ausgeführt wird. Jedoch ist das Sammeln von App-Daten von der Firmwareebene aus noch komplexer als von der Betriebssystemebene, auch wenn es dafür bereits erste technische Lösungen gibt.[36]

**V. Weitere Varianten**

Während bei aktuellen Implementierungen der QTKÜ intuitiv Kommunikationsdaten direkt auf dem Endgerät vor oder nach der Verschlüsselung erhoben werden, wurde kürzlich eine Variante vorgestellt, bei der stattdessen nur das Schlüsselmaterial ausgeleitet wird.[37] Hierbei werden die verschlüsselten Kommunikationsdaten weiterhin traditionell im Zuge einer TKÜ aufgezeichnet, während der Zugang zum Endgerät nur benutzt wird, um die für die Entschlüsselung notwendigen Schlüssel zu extrahieren. Ein großer Vorteil dieser Variante ist die garantierte Beschränkung der Datenerhebung auf Kommunikationsdaten, da durch die TKÜ garantiert ist, dass nur laufende Kommunikation erfasst wird.

Daneben gibt es noch technische Ansätze, die nicht nur auf ein und derselben Privilegienstufe operieren, sondern eine Kombination von Exploit und Herstellerkooperation vorschlagen.[38] Die Kernidee ist dabei, den Zugang und die Beweismittelerhebung technisch strikt zu trennen, sodass die Chance für Manipulation und Missbrauch minimiert, wobei gleichzeitig die Zuverlässigkeit erhöht wird. Der Gerätehersteller verhilft bei diesem Ansatz den Strafverfolgungsbehörden dabei zwar dazu, genug Privilegien für das Sammeln von Beweismitteln auf dem Endgerät zu erlangen, bringt aber gleichzeitig einen Mechanismus zur Beobachtung mit ein, der ein geschütztes technisches Protokoll über den Ermittlungsvorgang führt.

**D. Gesellschaftliche Akzeptanz staatlicher Eingriffe**

Die gesellschaftliche Akzeptanz staatlicher Eingriffe ist für eine erfolgreiche Regulierung von großer Bedeutung, da diese bei fehlender Akzeptanz am Widerstand der Bevölkerung scheitern kann.

Die Akzeptanz gegenüber staatlicher Überwachung hängt von verschiedenen Faktoren ab, die in der bisherigen Forschung unterschiedlich stark untersucht wurden. Zur Akzeptanz gegenüber den spezifischen Überwachungsmaßnahmen (Q)TKÜ oder der Chatkontrolle existiert unseres Wissens nach derzeit jedoch keine Forschung. Auf Grundlage der bisherigen

---

[36] *Busch/Nicolai/Fleischer/Rückert/Safferling/Freiling*, in: Goel/Gladyshev/Johnson/Pourzandi/Majumdar (Hrsg.), Digital Forensics and Cyber Crime, 2021, S. 23; *Schulze/Lindenmeier/Röckl/Freiling*, in: Schrittwieser/Ianni (Hrsg.), Proceedings of the 2024 Workshop on Research on offensive and defensive techniques in the context of Man At The End (MATE) attacks, 2024, S. 11.
[37] *Lindenmeier/Hammer/Gruber/Röckl/Freiling* Forensic Science International: Digital Investigation 50, 301796.
[38] *Lindenmeier/Gruber/Freiling* Digital Threats: Research and Practice 5 Ausgabe 3, 1.

Forschungsergebnisse, beispielsweise zur Akzeptanz der Vorratsdatenspeicherung lassen sich jedoch generelle Rückschlüsse zu den Faktoren ziehen, die gesellschaftliche Akzeptanz positiv oder negativ beeinflussen. Der Fokus soll dabei besonders auf den Einfluss der rechtlichen und technischen Ausgestaltung der Maßnahmen auf die Akzeptanz gelegt werden.

Grundsätzlich ist die Überwachung digitaler Kommunikation in Deutschland gesellschaftlich weniger akzeptiert als Überwachung in öffentlichen Bereichen, wie beispielsweise Videoüberwachung im öffentlichen Raum.[39]

Dabei sind gezielte Maßnahmen, die sich gegen konkrete Verdächtige richten, eher akzeptiert als generelle Maßnahmen, die potenziell die gesamte Bevölkerung betreffen.[40] Repräsentative Umfragen in Deutschland aus den Jahren 2009 und 2011 zeigen beispielsweise eine geringe Akzeptanz der Vorratsdatenspeicherung, also die generelle Speicherung von Telefon- und Internetverbindungsdaten durch Telekommunikationsanbieter. Rund zwei Drittel der Befragten lehnen diese Maßnahme ab. Die unbemerkte Online-Durchsuchung von persönlichen Computern verdächtiger Personen wird vergleichsweise stärker akzeptiert, nur rund die Hälfte der Befragten lehnt diese Maßnahme ab.[41]

Darüber hinaus zeigen Studien, dass die wahrgenommene Sicherheit im Umgang mit den erhobenen Daten eine Rolle bei der Akzeptanz der Maßnahme spielt.[42] Dieses Ergebnis zeigt uns, dass bisher ungeklärte Fragen zur postulierten Maßnahme der Herstellerkooperation die gesellschaftliche Akzeptanz beeinflussen könnten. Wie werden die erhobenen Daten gesichert und verarbeitet? Und wie hoch ist das gesellschaftliche Vertrauen in diese nicht staatlichen Akteure, die Daten verlässlich zu schützen? Diese ungeklärten Fragen zeigen den Bedarf nach weiterer Forschung, die Einblicke in das Zusammenspiel von technischer und rechtlicher Ausgestaltung von Eingriffen liefert.

Neben der Ausgestaltung der Eingriffe beeinflussen allerdings auch individuelle Faktoren maßgeblich die Akzeptanz staatlicher Eingriffe. Dazu zählen die Furcht vor Kriminalität und Terrorismus,[43] das Vertrauen in den Staat und seine Institutionen[44] und allgemeine politische Einstellungen.

Seit den letzten großen Erhebungen zur Akzeptanz staatlicher Überwachung, die teils mehr als zehn Jahre zurückliegen, haben jedoch erhebliche gesellschaftliche Veränderungen stattgefunden. Nicht zuletzt sind digitale Dienste immer tiefer in den Alltag weiter Teile der Gesellschaft vorgedrungen und aus vielen Bereichen nicht mehr wegzudenken. Dadurch wächst auch die Gefahr, dass der Kernbereich der privaten Lebensführung durch staatliche Maßnahmen betroffen sein könnte.[45] Auch die jüngsten Debatten um die Einführung der Chatkontrolle zeigen ein hohes gesellschaftliches Bewusstsein für mögliche Gefahren. Dies könnte zu einer geringeren Akzeptanz staatlicher Maßnahmen in der Bevölkerung führen.

---

[39] *Antoine* Surveillance & Society 21, 409; *Jäger* Frontiers in Political Science 4, 1006711.

[40] *Ziller/Helbling* European Journal of Political Research 60, 994; *Antoine* Surveillance & Society 21, 409; *Jäger* Frontiers in Political Science, 4, 1006711.

[41] *Lüdemann/Schlepper*, in: Zurawski (Hrsg.), Überwachungspraxen. Praktiken der Überwachung und Kontrolle, 2011, S. 119; *Bug/Bukow* German Politics 26, 292; *Trüdinger/Steckermeier* Government Information Quarterly 34, 421.

[42] *Ziller/Helbling* European Journal of Political Research 60, 994.

[43] *Bali* Policy Studies Journal 37, 233; *Davis/Silver* American journal of political science 48, 28; *Garcia/Geva* Terrorism and Political Violence 28, 30; *Trüdinger/Steckermeier* Government Information Quarterly 34, 421; *Ziller/Helbling* European Journal of Political Research 60, 994.

[44] *Bali* Policy Studies Journal 37, 233; *Davis/Silver* American journal of political science 48, 28; *Trüdinger/Steckermeier* Government Information Quarterly 34, 421.

Gleichzeitig hat jedoch auch die Angst vor Kriminalität in Deutschland stark zugenommen, obwohl die Zahlen der polizeilichen Kriminalstatistik auf einen Rückgang der erfassten Straftaten hindeuten.[46] Da die Kriminalitätsfurcht in direktem Zusammenhang mit der Akzeptanz staatlicher Maßnahmen steht, könnte auch die gesellschaftliche Akzeptanz gegenüber diesen Maßnahmen gestiegen sein.

Aufgrund dieser und weiterer gesellschaftlicher Veränderungen wie der sich aktuell zuspitzenden geopolitischen Weltlage sind Studien erforderlich, die aktuelle Daten zur gesellschaftlichen Akzeptanz staatlicher Eingriffe liefern und darüber die Wechselwirkungen mit der rechtlichen und technischen Ausgestaltung der Eingriffe auf die Akzeptanz im Blick behalten.

**E. Fazit: Problembereiche der Regulierung**

Der Zugriff auf die Inhalte von Ende-zu-Ende-verschlüsselten Kommunikationsverbindungen ist ein wiederholt vorgebrachtes Desiderat der Strafverfolgung, das durch die aktuellen Regelungen zur QTKÜ in der StPO ohne Mitwirkungspflichten Dritter reguliert wurde. Die polarisierte Diskussion zur Chatkontrolle hat die Frage von Mitwirkungspflichten wieder aufgeworfen. Vor dem Hintergrund der geringen medialen Aufmerksamkeit gegenüber Mitwirkungspflichten im Bereich der klassischen TKÜ erscheint diese Situation geradezu paradox. Ausgangspunkt dieses Beitrages war somit die Frage, warum der Gesetzgeber nicht schon längst die Hersteller von Kommunikationssoftware zur Bereitstellung von Abhörschnittstellen verpflichtet hat? Wie unser Beitrag zeigt, liegt dies nicht nur am mangelnden politischen Willen, sondern auch an der Komplexität der Regulierungsoptionen und ihrer Problembereiche.

Der erste und augenfälligste Problembereich ist: *Wer* wird verpflichtet? Es gibt im Bereich der Ende-zu-Ende-Verschlüsselung viele Anbieter, deren Software auf unterschiedlichen Privilegienstufen operiert. Technisch macht es zwar Sinn, auf höchster Privilegienstufe zu agieren und somit Konzerne wie Apple, Google und Samsung zu verpflichten. Dieses Vorhaben erscheint jedoch ohne größere internationale Koalitionen politisch aussichtslos, und zwar unabhängig von der zur Regulierung gewählten Privilegienstufe. Der Regulierungsvorschlag zur Chatkontrolle, der eine solche Verpflichtung zum Ziel hat, kann als politischer „Testballon" in dieser Richtung verstanden werden. Die handwerkliche Durchführung wirkt dabei aber ähnlich pedantisch wie die ersten Regulierungsversuche zur

---

[45] Zum Kernbereichsschutz bei TKÜ z.B. BVerfGE 113, 348 (390 ff.); Zur Möglichkeit umfassender Erkenntnisse über das Privatleben und einer Profilbildung BVerfG BeckRS 2025, 19413 - *Trojaner II*, Rn. 188 ff.
[46] *Bindler* DIW aktuell 108, S. 5.

QTKÜ in Deutschland in Form des Verfassungsschutzgesetzes von Nordrhein-Westfalen vom 20.12.2006, das Gegenstand des Urteils zur Online-Durchsuchung des Bundesverfassungsgerichts war.[47]

Die daran anknüpfende Frage lautet: *Wie* verpflichtet man die Softwarehersteller? Dies betrifft den Problembereich der technischen Realisierungsmöglichkeiten eines Zugriffs auf Endgeräte, die einen Ausgleich finden zwischen den legitimen Wünschen der Strafverfolgungsbehörden nach „einfachem" Zugriff auf die Inhalte Ende-zu-Ende-verschlüsselter Kommunikation und dem wirksamen Schutz der Bürger vor automatisierter Massenüberwachung. Es ist darum notwendig, die gerichtliche Kontrolle, die in der aktuellen Fassung der §§ 100a, 100e StPO bereits ausgestaltet ist, durch eine weitere Kontrollkomponente zu erweitern.

Genau hier kommen die Softwarehersteller ins Spiel. Würde man sie dazu verpflichten, abhängig vom Standort des Endgerätes (deutsches Staatsgebiet) und innerhalb einer Frist (Dauer des Beschlusses) Inhalte auszuleiten, käme ihnen genau die Rolle zu, die bisher Telekommunikationsdienstleister beim Zugriff auf unverschlüsselte Inhalte spielten.[48] Wie der Verweis auf die technische Literatur zeigt, wäre der Zugriff auf unterschiedlichen Privilegienstufen realisierbar. Durch die Mitwirkungspflicht der Softwarehersteller entstünde gleichzeitig eine Mitwirkungsnotwendigkeit, also eine Kooperationsabhängigkeit, die sich moderierend auf überbordende Bedürfnisse des Staates und damit positiv auf die gesellschaftliche Akzeptanz auswirken könnte.

Zuletzt stellt sich die Frage: Ist eine solche Verpflichtung politisch-gesellschaftlich durchsetzbar? Die Antwort auf diese Frage hängt kritisch mit der gesellschaftlichen Akzeptanz der Ausgestaltung der technischen Maßnahmen zusammen, ein generelles Problem der Technikregulierung im Sicherheitsbereich. Die technischen Realisierungsalternativen liegen – wie oben beschrieben – mittlerweile vor. Der gesellschaftliche Diskurs und damit die demokratische Willensbildung zu einer technisch derart komplexen Materie hingegen hat gerade erst begonnen.

---

[47] BVerfGE 120, 274.

[48] Dabei könnten die Risiken, welche sich durch den Einbezug weiterer Beteiligter in den Abhörprozess ergeben, durch eine Kontrolle analog zu der Kontrolle durch die Bundesnetzagentur im Rahmen der klassischen TKÜ eingedämmt werden. Zu Risiken siehe BVerfG BeckRS 2025, 19412 – *Trojaner I*, Rn. 114.